\documentclass{PoS}
\def\MET{\ensuremath{E_{\mathrm{T}}^{\mathrm{miss}}}}
\def\met{\ensuremath{E_{\mathrm{T}}^{\mathrm{miss}}}}
\def\pT{\ensuremath{p_{\mathrm{T}}}}
\def\pt{\ensuremath{p_{\mathrm{T}}}}
\def\antibar#1{\ensuremath{#1\bar{#1}}}
\def\ttbar{\antibar{t}}

\def\tev{\ifmmode {\mathrm{\ Te\kern -0.1em V}}\else
                   \textrm{Te\kern -0.1em V}\fi}%
\def\gev{\ifmmode {\mathrm{\ Ge\kern -0.1em V}}\else
                   \textrm{Ge\kern -0.1em V}\fi}%
\def\mev{\ifmmode {\mathrm{\ Me\kern -0.1em V}}\else
                   \textrm{Me\kern -0.1em V}\fi}%

\title{Search for FCNC in top-quark events in ATLAS}
\ShortTitle{Search for FCNC in top-quark events in ATLAS}
\author{\speaker{Markus Cristinziani}\thanks{Supported by DFG grant CR-312/1-2}\\ 
        {\rm On behalf of the ATLAS Collaboration}\\
        University of Bonn and CERN\\
        E-mail: \email{cristinz@uni-bonn.de}}

\abstract{
Searches for flavor changing neutral current (FCNC) processes in
top-quark production and decays by the ATLAS Collaboration are presented. Data
collected from $pp$ collisions at the LHC at a centre-of-mass energy of
$\sqrt{s}=7$~\tev\ during 2011, corresponding to an integrated
luminosity of 2.05~fb$^{-1}$, are used. 
In a first analysis single top-quarks produced via FCNC 
are searched for.  Candidate events with a semileptonic top-quark 
decay signature are classified as signal or background-like events by using
several kinematic variables as input to a neural network. No signal is observed
in the neural network output distribution and a Bayesian upper limit is placed 
on the production cross-section. The observed upper limit is converted using a 
model-independent approach into upper limits on the coupling strengths 
$\kappa_{ugt}/\Lambda < 6.9\cdot 10^{-3}$~TeV$^{-1}$ and 
$\kappa_{cgt}/\Lambda < 1.6\cdot 10^{-2}$~TeV$^{-1}$, where $\Lambda$ is the 
new physics scale, and on the branching fractions 
${\cal B}(t \rightarrow ug) < 5.7\cdot 10^{-5}$ and 
${\cal B}(t \rightarrow cg) < 2.7\cdot 10^{-4}$.
A second search is performed for top-quark pair-production events, with 
one top quark decaying through the $t\to Zq$ FCNC ($q=u,c$) channel, and 
the other through the Standard Model dominant mode $t\to Wb$. 
Only the decays of the $Z$ boson to charged leptons and leptonic 
$W$-boson decays are considered as signal. 
No evidence for an FCNC signal is found and thus an upper limit
on the $t\to Zq$ branching ratio of BR$(t\to Zq)<0.73$\% is set at
the 95\% confidence level.
}

\FullConference{36th International Conference on High Energy Physics\\
                 4-11 July 2012\\
                 Melbourne, Australia}

\begin{document}

\section{Flavor-changing neutral currents and the top quark}

In the Standard Model (SM), flavor-changing neutral current (FCNC) 
processes are forbidden at tree level and highly suppressed (${\cal O}(10^{-14})$)
at higher orders due to the GIM mechanism~\cite{GIM}.
Extensions of the SM with new sources of flavor predict 
higher rates for FCNCs involving the top quark; these extensions include new exotic 
quarks, new scalars, supersymmetry, or technicolor 
(for a review see Ref.~\cite{Agu}).
The top-quark FCNC decay BR in these models is enhanced dramatically 
and can be as high as $\sim 2\times 10^{-4}$ in certain $R$-parity violating SUSY models.

FCNC top-quark decays can be studied directly by searching for final 
states with the corresponding decay particles.
The $t \rightarrow qg$ mode, with $q = u,c$, 
is almost impossible to separate from multijet-production and 
a better sensitivity can be achieved in the search for anomalous single top-quark production.
Here, a quark and a gluon coming from the colliding protons interact 
to produce a single top-quark. The most general effective Lagrangian ${\cal L}_{\rm eff}$ for this process 
can be written as:
\begin{equation}
{\cal L}_{\rm eff}=g_{s} \sum_{q=u,c} \frac{\kappa_{qgt}}{\Lambda} \bar{t}\sigma^{\mu\nu}T^{a}(f^{L}_q P_L+f^R_q P_R)qG^{a}_{\mu\nu} + h.c. , \nonumber
\end{equation}
where the $\kappa_{qgt}$ are dimensionless parameters that relate the strength 
of the new coupling to the strong coupling constant $g_{s}$ and ${\Lambda}$ 
is the new physics scale, related to the mass cutoff scale above which the effective theory breaks down.

The analysis~\cite{ugt} is performed using $\sqrt{s}=7$~TeV $pp$-collision data 
recorded by ATLAS~\cite{ATL} in 2011 and corresponding
to a total integrated luminosity of $2.05\pm0.08$~fb$^{-1}$.

Electron candidates are required to satisfy \pT\ $> 25$ GeV and 
$\mathrm{|\eta_{\rm clus}| <}$ 2.47, where $\mathrm{\eta_{\rm clus}}$ is the
pseudorapidity of the cluster of energy deposits in the calorimeter. 
A veto is placed on candidates in the calorimeter barrel-endcap transition region, 
$1.37 < |\eta_{\rm clus}| < 1.52$. 
High-\pT~electrons associated with the $W$-boson decay can be mimicked by hadronic jets reconstructed as electrons,
electrons from decays of heavy quarks, and photon conversions.  
These backgrounds are suppressed via isolation 
criteria in a cone around the electrons.
Muon candidates are required to have a transverse momentum 
\pT $> 25$ GeV and to be in the pseudorapidity region of $|\eta|<2.5$.
Isolation criteria are applied to reduce background events in which a high-$\pT$ muon is produced
in the decay of a heavy quark.  
Jets are reconstructed using the anti-$k_{t}$ algorithm with the distance parameter $R$ set to 0.4, 
and exactly one reconstructed jet with $\pT>25$~GeV is required. 
Due to the presence of a neutrino, a missing transverse momentum
$\MET > 25$~GeV is required. 
Multijet background events are further suppressed by combined requirements on $\MET$ and 
the reconstructed $W$-boson transverse mass.
Candidate events are required to have exactly one lepton and one jet. The selected jet has to be  
identified as stemming from a $b$-quark ($b$-tagged).

Multijet events may be selected if a jet is misidentified as an isolated lepton or 
if the event has a non-prompt lepton that appears isolated.
A binned maximum-likelihood fit to the $\MET$ distribution is used to estimate the multijet background normalization. 
A template of the multijet background is modelled using electron-like jets selected from
jet-triggered collision data.
The uncertainty in the multijet background normalization is estimated to be 50\%.

\begin{figure}[htbp]
\begin{center}
\includegraphics[width=0.49\textwidth]{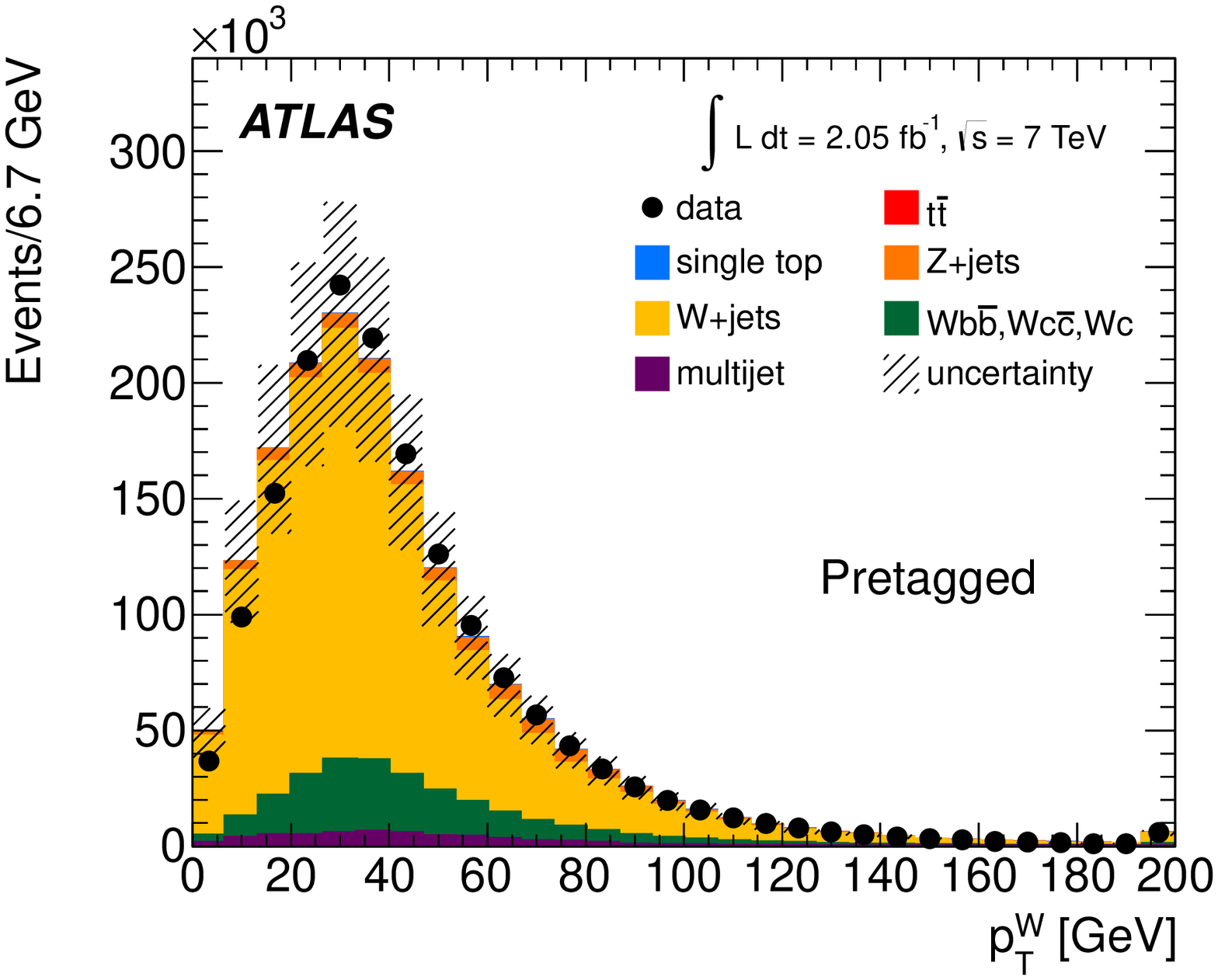}
\includegraphics[width=0.49\textwidth]{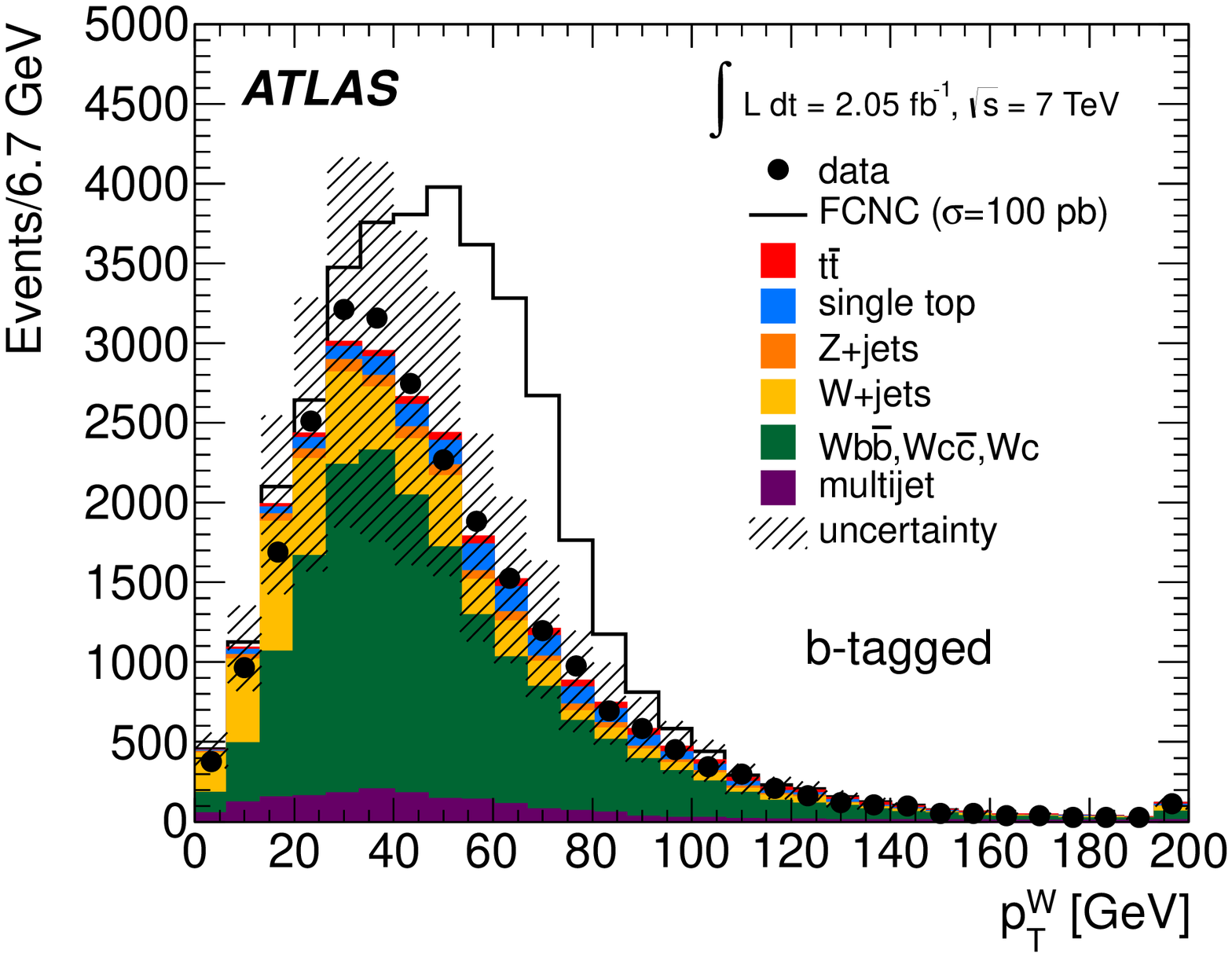}
\caption{\label{fig:Wpt}
Data and expected distributions of the transverse momentum of the $W$ boson,
for the selection before (left) and after (right) $b$-tagging, for the 
electron and muon channel combined~\cite{ugt}.}
\end{center}
\end{figure}

In the final sample, $26,\!223$ events are observed in data compared to a SM expectation
of $24,\!000 \pm 7,\!000$ events, dominated by $Wc$+jets events ($12,\!100 \pm 6,\!700$). Figure~\ref{fig:Wpt} shows
good agreement between the data and expectation.
Given the large uncertainty in the background and the small number of expected signal events, 
a neural-network classifier is used to separate signal events from background events.

The $qg\to t\to b\ell\nu$ process is characterized by three main differences from SM processes that pass the event selection cuts:
(1) The top quark is produced almost without transverse momentum. 
(2) The $W$ boson from the top-quark decay has a very high momentum. 
(3) The FCNC processes are predicted to produce four times more single top quarks than anti-top quarks, while this is at most two in SM processes.

\begin{figure}[htbp]
\centering
\includegraphics[width=0.49\textwidth]{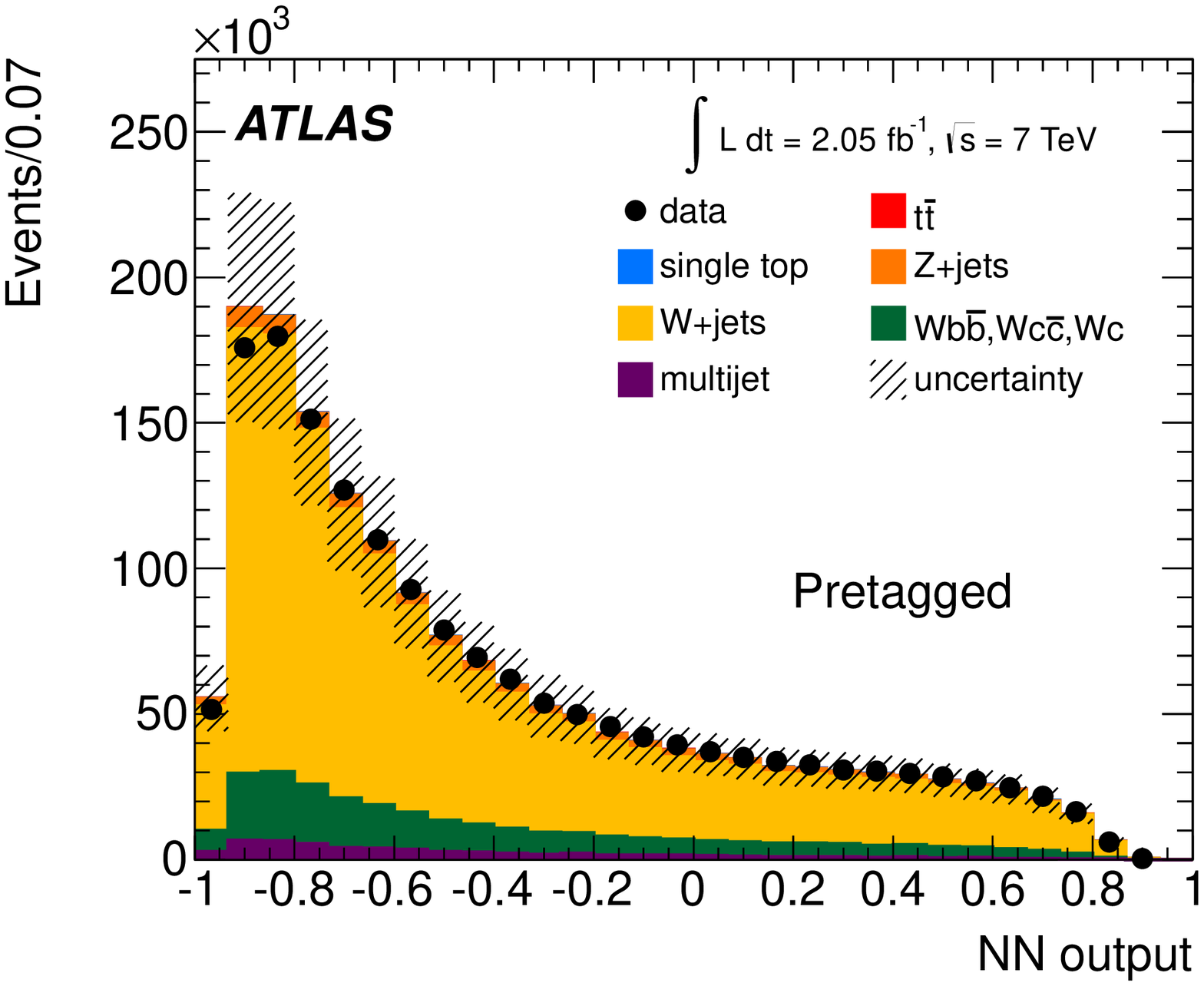}
\includegraphics[width=0.49\textwidth]{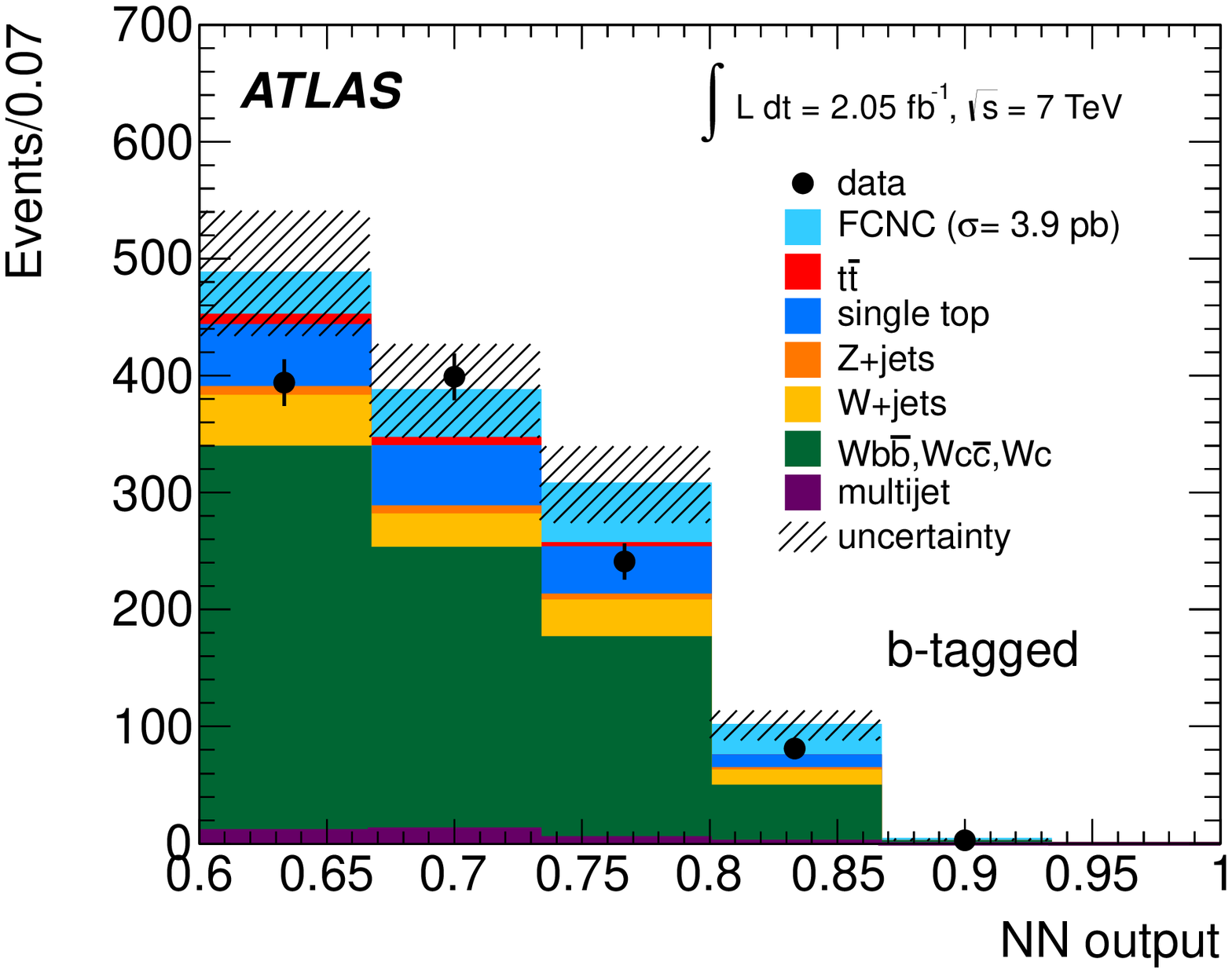}\\
\caption{\label{fig:nn}
Neural network output distribution in the sample before requiring $b$-tagging (left) and in the final
sample, zoomed into the signal region (right). The FCNC single top quark process is normalized to the observed limit of 3.9~pb~\cite{ugt}.}
\end{figure}

Eleven variables are selected as input to the neural network. The three most significant variables are: 
the $p_{\mathrm{T}}$ of the $W$ boson,
the $\Delta R$ between the $b$-tagged jet and the lepton, 
and the charge of the lepton.
The resulting neural network output distribution for the various processes, 
is shown in Figure~\ref{fig:nn}. 
Signal-like events have output values close to 1.
Good agreement between the neural network output distributions for data and simulated events is found.

Systematic uncertainties affect the signal acceptance, the normalization of the individual backgrounds, 
and the shape of the neural network output distributions. They lead to uncertainties in the rate estimation 
as well as distortions of the neural network output distribution and are implemented as such in the 
statistical analysis.  Several sources of systematic uncertainties have been considered. 
The three dominant uncertainties are: (1)
The uncertainty in the jet energy scale, varying between 2.5\% and 8\% (3.5\% and
14\%) in the central (forward) region, depending on jet $p_{\mathrm{T}}$ and
$\eta$, including gluon-fraction uncertainties  and mis-measurements due to close-by jets. 
An additional jet energy scale uncertainty of up to 2.5\% is applied for $b$-quark jets. 
(2) The amount of initial and final state radiation, which is
varied by modifying parameters in {\sc PYTHIA} in a range comparable to those used in the Perugia Soft/Hard tune
variations.
(3) The $b$-tagging efficiencies and mis-tag rates, 
having uncertainties from 8--16\% and 23--45\%, respectively.

A Bayesian statistical analysis using a binned likelihood method applied to 
the neural network output distributions for the electron and muon channel is performed to measure or set 
an upper limit on the FCNC single top-quark production cross-section.

\section{Search for FCNC in top-quark decays $t \to Zq$}

In a second analysis~\cite{tZq} a search is performed for FCNC top-quark decays in $\ttbar$
events, where one of the top quarks decays to a $Z$ boson and a quark, $t\rightarrow Zq$,
while the other decays through the SM $t\rightarrow Wb$ channel. 
Only leptonic decays of the $Z$ and $W$
bosons are considered, yielding a final-state topology
characterized by the presence of three isolated charged leptons, at
least two jets, and significant $\MET$. 

The selection of leptons, jets, and $\MET$ is close to that
used for the ATLAS measurement of the \ttbar\ production cross
section in the dilepton channel~\cite{DIL}.  
Events are selected with either three well-identified leptons with
$\pt > 20 \gev$ (3ID) or two well-identified leptons with $\pt>20 \gev$ 
and an additional track-lepton with $\pt>25 \gev$, which fails lepton
identification (2ID+TL). In the following only the 3ID analysis is 
presented, since the background estimation technique for the 2ID+TL channel is 
described in~\cite{DIL} and the remaining analysis follows the main analysis.

\begin{figure}
\begin{center}
\includegraphics[width=0.59\textwidth]{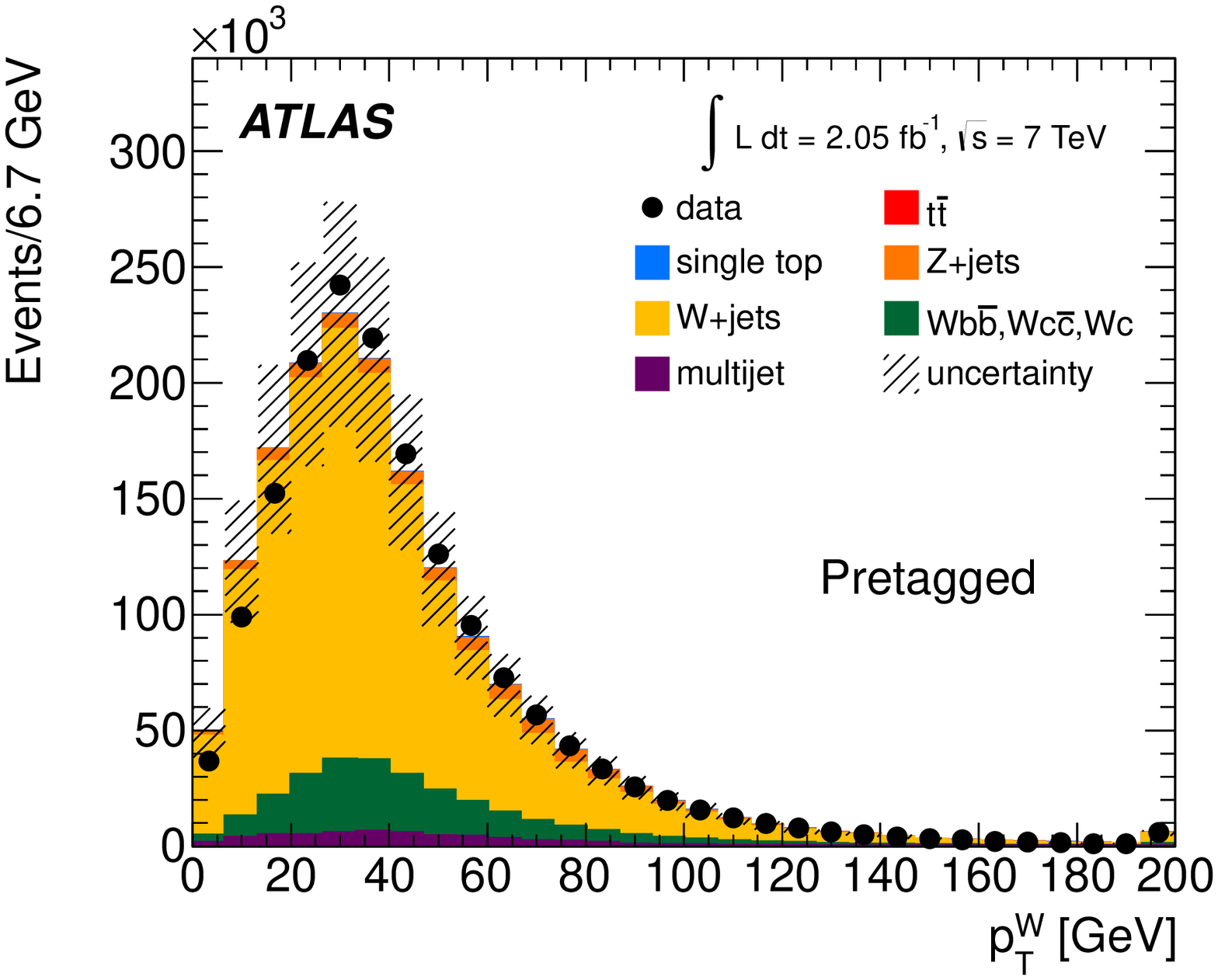}
\includegraphics[width=0.39\textwidth]{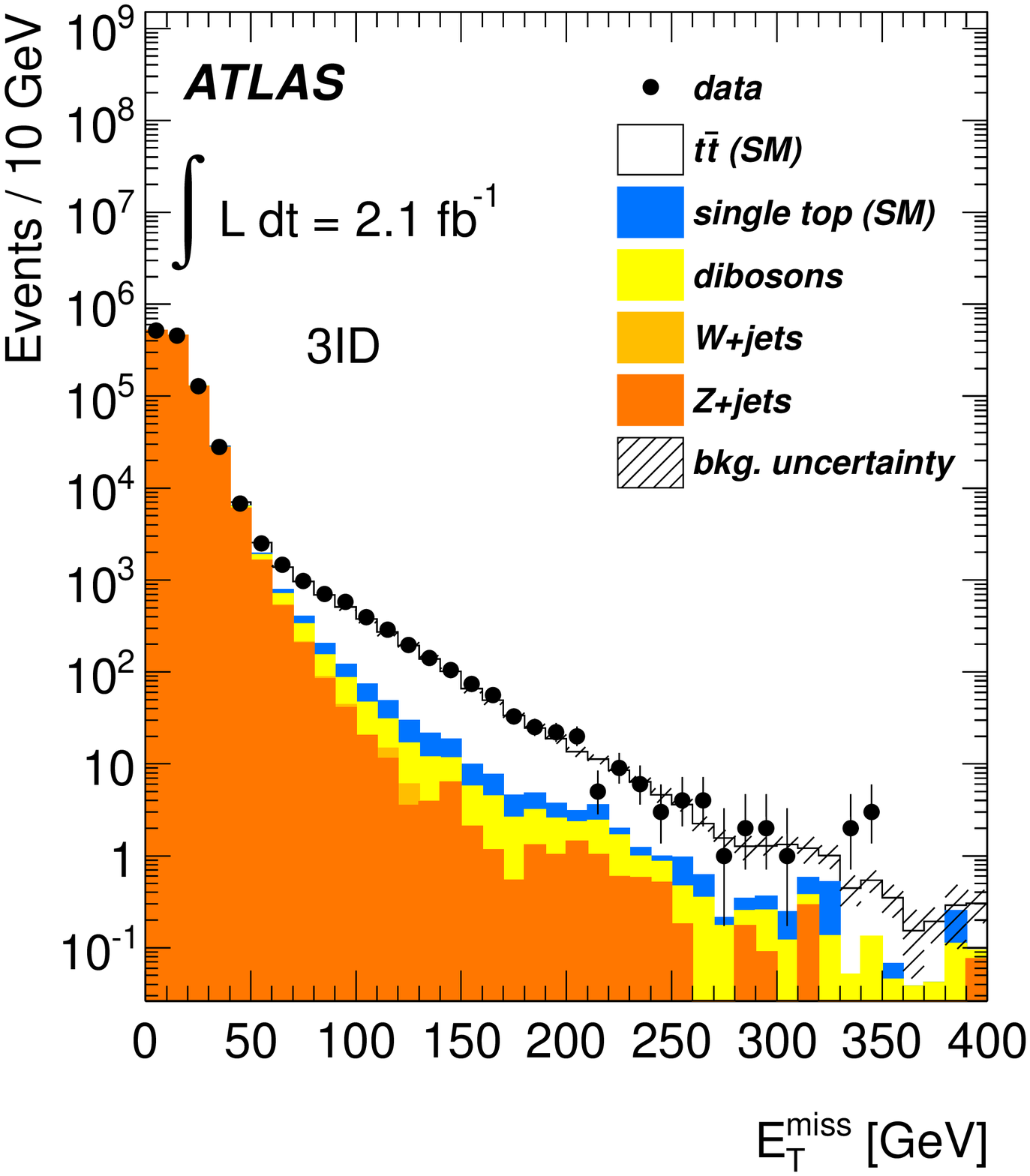}
\caption{\label{fig:met}\met distribution
for events with three identified leptons (at least one
OSSF, consistent with a $Z$-boson), 
but no jets or $\met$ requirement (left) and 
for events in a control region, defined by 
two same-flavor, OS leptons, consistent with a $Z$-boson (right).
The $\ttbar\rightarrow WbZq$ distribution is
normalized to the observed limit.~\cite{tZq}
}
\end{center}
\end{figure}

Events are required to have a same-flavor, opposite-sign (OS) lepton pair 
with an invariant mass within 15~\gev\ of $m_Z$.  
In addition, signal candidates are required to have at 
least two jets and $\met >$ 20 \gev. 
Figure~\ref{fig:met} (left) shows the $\met$ distribution for 
events prior to the final selection requirements. 
These are events with three identified leptons with at
least one opposite-sign, same-flavor pair (OSSF) with the $m_Z$ invariant mass
cut applied, but no jets or $\met$ requirement.  

Selected events are required to be kinematically consistent with
$\ttbar\rightarrow WbZq$ through a $\chi^2$ minimized with respect
to jet (the two highest-$\pt$) and lepton assignments and the longitudinal momentum of the
neutrino, $p_z^\nu$. The widths are determined from simulation.
All jet and lepton assignments are tried,
subject to the requirement that the $Z$ candidate be built from OSSF leptons.
From all combinations, the one with the smallest $\chi^2$ is chosen along
with the corresponding $p_z^\nu$ value. Events are required 
to have reconstructed $m_t$, $m_W$, $m_Z$ within respectively 40~\gev, 30~\gev, and 15~\gev,
of the nominal mass. The signal efficiency for 
$\ttbar\rightarrow WbZq$, after all
selection requirements, is determined to be $(0.205 \pm 0.025)\%$.

Backgrounds can be divided into two categories: those
with three real leptons and those with at least one fake lepton.
Backgrounds with three real leptons arise from diboson ($WZ$ and
$ZZ$) production with additional jets, and are evaluated using simulation.
The background in which exactly one of the leptons
is a fake lepton, is evaluated using a combination of data and MC
samples. The dominant contribution in this category comes from
$Z$+jets events, with a leptonic $Z$ decay, in which one of the jets
is misidentified as a third lepton.  To evaluate this background a
data-driven (DD) method is used with control regions (CR) in the
($\met$, $m_{\ell \ell}$) plane, by selecting events with exactly 
two OS leptons (no third lepton is allowed)
and $|m_Z-m_{\ell\ell}|<15$~\gev\ in six
different $\met$ bins from 0 GeV to $\geq$50 GeV. 
For each $\met$ bin considered, the
background-subtracted data/simulation ratio in the CR is applied to
the simulated $Z+$jets background in the signal regions, in
order to evaluate the expected number of $Z+$jets events in
data. Due to the small MC event sample after the final selection, 
the $Z+$jets background is evaluated using a loosened lepton selection
and a multiplicative rejection factor of $0.063 \pm 0.013$ to account
for the loosened selection. 
The remaining backgrounds with one fake lepton (dileptonic $t\bar t$, 
$Wt$-channel single-top and $WW$ production) are evaluated using 
simulated samples.
The contribution from multi-jet, $W$+jets, single-top and $t\bar t$
single-lepton decay events, where two or three jets are reconstructed
as leptons, is estimated with a DD technique, making use of events
with same-charge leptons. No data event passed the selection after 
requiring three leptons with the same charge.

\begin{figure}
\begin{center}
\includegraphics[width=0.49\textwidth]{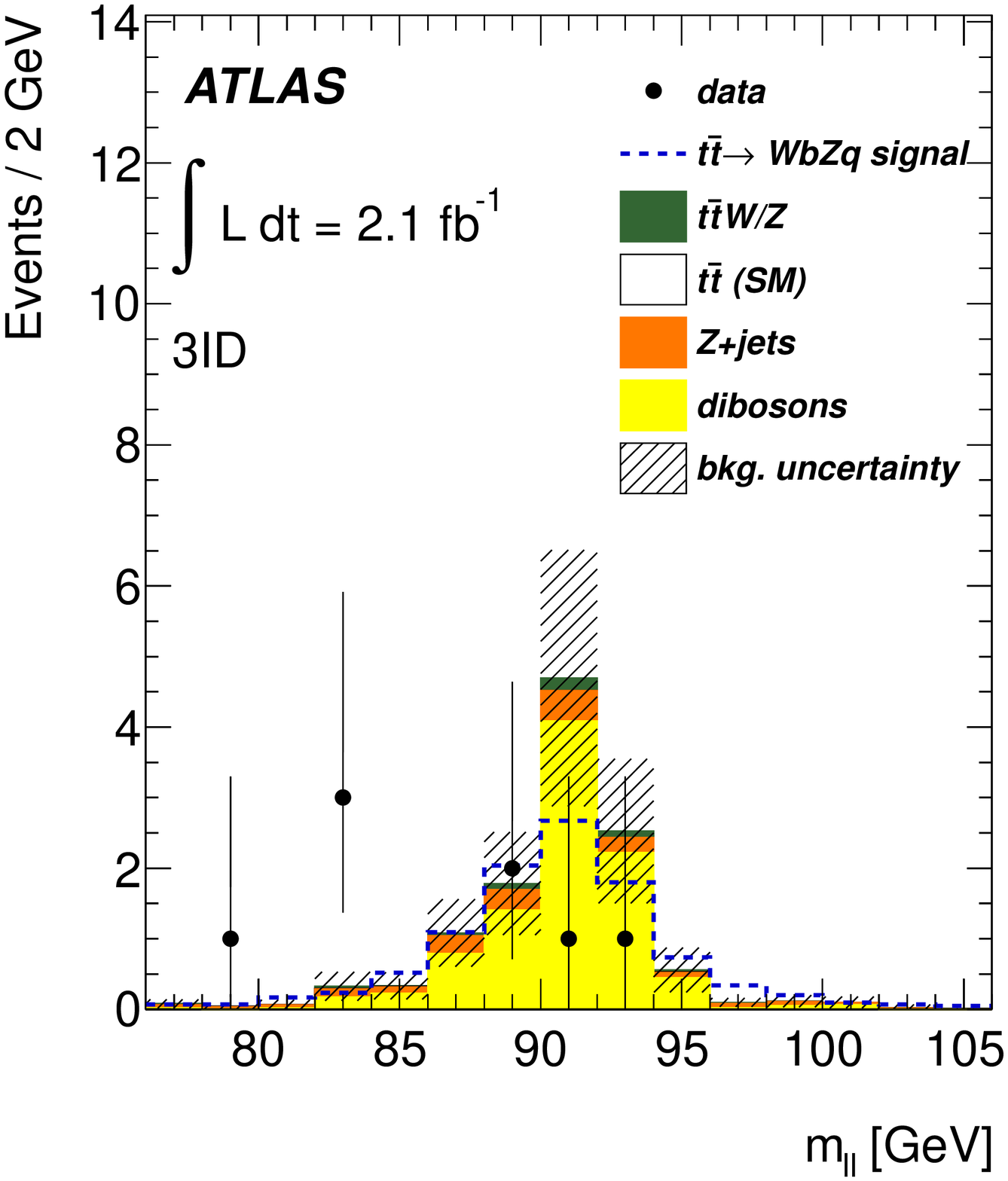}
\includegraphics[width=0.49\textwidth]{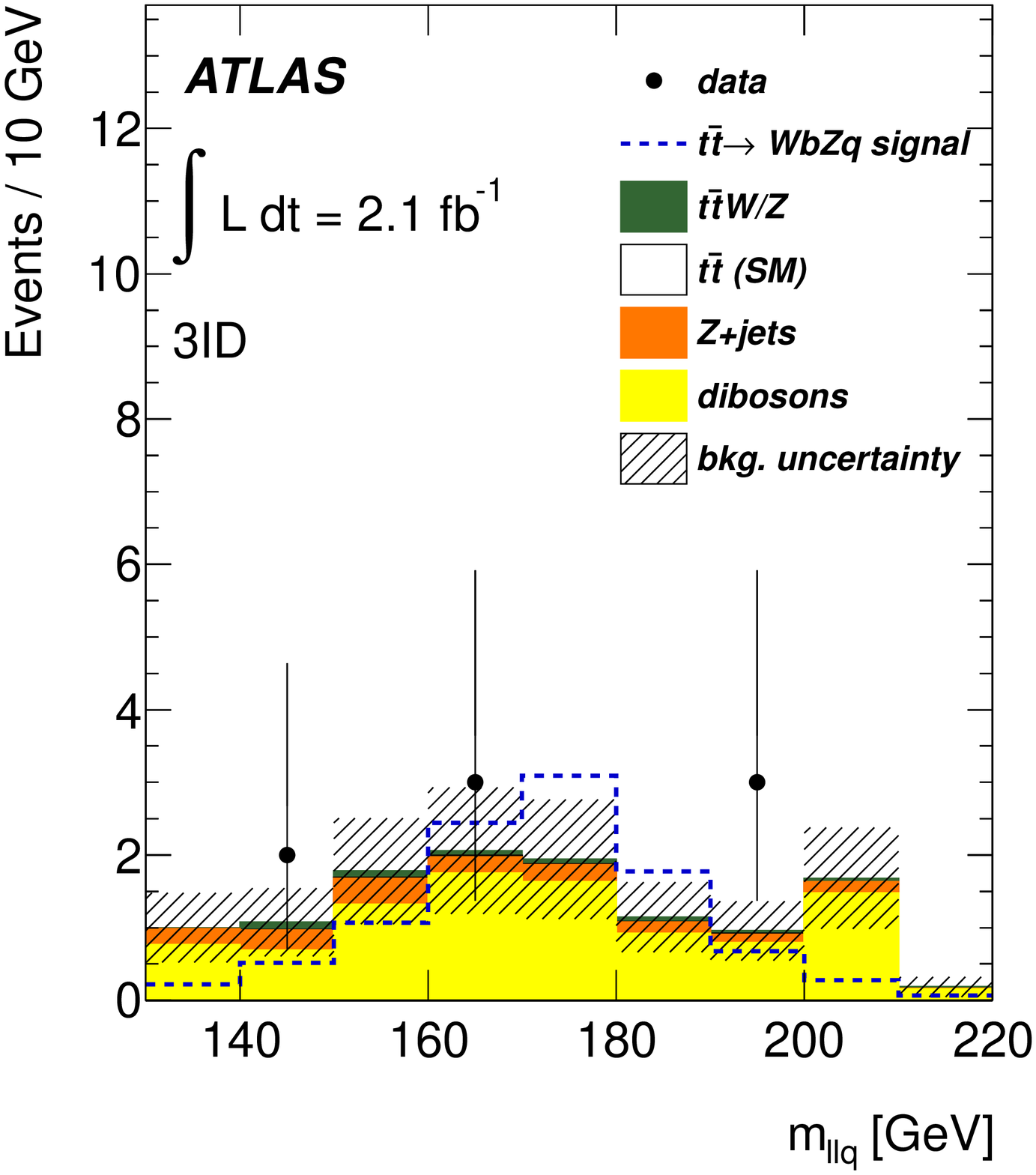}
\caption{\label{fig:kinsum}$Z$-boson (left) and top-quark
mass distributions (right) for the FCNC decay hypothesis
after all selection requirements. The $\ttbar\rightarrow WbZq$ distributions are
normalized to the observed limit in each channel~\cite{tZq}.}
\end{center}
\end{figure}

The total number of expected background events is
$11.8 \pm 4.4$, dominated by diboson events ($9.5 \pm 4.4$). 
Figure~\ref{fig:met} (right) shows good agreement 
in the $\met$ distribution of data and expected backgrounds in 
background-dominated control region.
Figure~\ref{fig:kinsum} shows the reconstructed candidate $Z$-boson 
and top-quark masses for the 
FCNC decay hypothesis in the selected candidate events, compared 
with the expectations from SM backgrounds and the FCNC signal.

A number of systematic uncertainties can influence the expected
number of signal and/or background events. The effect of each source
of systematic uncertainty is studied by independently varying the
corresponding central value by the estimated uncertainty. For each
variation, the total number of expected background events and the signal
efficiencies are compared with the reference values.
The dominant source of systematic uncertainty is 
the $ZZ$ and $WZ$ simulation modeling, which is
estimated using the Berends-Giele scaling with an uncertainty of 24\% 
per jet, added in quadrature. An uncertainty
of 4\% is included for the 0-jet bin. The $ZZ$ and $WZ$ cross sections are
varied by their theoretical uncertainty of 5\%. 
The dominant uncertainty in the 2ID+TL channel is the systematic
uncertainty on the fake-TL prediction.

\section{Results}

A data sample selected to consist of events with an isolated electron or muon, 
missing transverse momentum and a $b$-quark jet 
has been used to search for FCNC production of single top-quarks at the LHC.
No evidence for such processes is found and the upper limit at 95\% C.L. 
on the production cross-section is 3.9 pb.
The limit is converted into limits on the coupling constants
$\kappa_{ugt}/\Lambda < 6.9\cdot 10^{-3}$~TeV$^{-1}$ and 
$\kappa_{cgt}/\Lambda < 1.6\cdot 10^{-2}$~TeV$^{-1}$, where $\Lambda$ is the 
new physics scale and on the branching fractions 
${\cal B}(t \rightarrow ug) < 5.7\cdot 10^{-5}$ assuming ${\cal B}(t \rightarrow cg)=0$, 
and ${\cal B}(t \rightarrow cg) < 2.7\cdot 10^{-4}$ assuming ${\cal B}(t \rightarrow ug)=0$

The search for the $t\to Zq$ decay mode is performed by studying top-quark 
pair production with one top quark decaying according to the SM 
and the other according to the FCNC ($t\bar t\to WbZq$).
No evidence for the $t\to Zq$ decay mode is found and a 95\% C.L. upper 
limits on the number of signal events are derived using the modified 
frequentist (CL$_{\mathrm s}$) likelihood method.
This results in BR$(t\to Zq)<0.73$\%, assuming BR$(t\to Wb)+$BR$(t\to Zq)=1$.

\end{document}